\documentclass[11pt,twoside]{article}
\usepackage{asp2010}

\usepackage{}
\resetcounters

\usepackage{epsf}
\begin{document}

\title{Shocks and Ejecta Mass: Radio Observations of Nova V1723 Aql}
\author{Jennifer H. S. Weston,$^1$ Jennifer L. Sokoloski,$^1$ Yong Zheng,$^1$ Laura Chomiuk,$^{2,3}$ Amy Mioduszewski,$^3$ Koji Mukai,$^{4,5}$ Michael P. Rupen$^{3}$, Miriam I. Krauss,$^3$ Nirupam Roy,$^3$ and Thomas Nelson.$^6$
\affil{$^1$ Columbia Astrophysics Laboratory, Columbia University, New York, NY 10027, USA}
\affil{$^2$ Department of Physics and Astronomy, Michigan State University,
	East Lansing, MI 48824, USA}
\affil{$^3$ National Radio Astronomy Observatory, P.O. Box O, Socorro, NM 87801, USA}
\affil{$^4$CRESST and X-ray Astrophysics Laboratory, NASA/GSFC, Greenbelt,
	MD 20771, USA}
\affil{$^5$Department of Physics, University of Maryland, Baltimore County,
	1000 Hilltop Circle, Baltimore, MD 21250, USA}
\affil{$^6$ School of Physics and Astronomy, University of Minnesota,
	116 Church Street SE, Minneapolis, MN 55455, USA}}

\begin{abstract}
The radio light curves of novae rise and fall over the course of months to years, allowing for detailed observations of the evolution of the nova shell. However, the main parameter determined by radio models of nova explosions --- the mass of the ejecta --- often seems to exceed theoretical expectations by an order of magnitude. With the recent technological improvements on the Karl G. Jansky Very Large Array (VLA), new observations can test the assumptions upon which ejecta mass estimates are based. Early observations of the classical nova V1723 Aql showed an unexpectedly rapid rise in radio flux density and a distinct  bump in the radio light curve on the rise to radio maximum, which is inconsistent with the simple model of spherical ejecta expelled in a single discrete event. This initial bump appears to indicate the presence of shocked material in the outer region of the ejected shell, with the emission from the shocks fading over time. We explore possible origins for this emission and its relation to the mass loss history of the nova. The evolution of the radio spectrum also reveals the density profile, the mass of the ejected shell, and other properties of the ejecta. These observations comprise one of the most complete, longterm set of multi-wavelength radio observations for any classical nova to date.
\end{abstract}

\section*{Introduction}
The nova V1723 Aql went into outburst on September 11th, 2010 \citep{Yam10, Balam10}
. Based solely on optical observations, V1723 Aql is a member of the ``fast nova" speed class \citep{Payne57},
fading by 2 magnitudes over 20 days, with emission lines implying an expansion velocity of $\sim$1500 km/s on the day of discovery \citep{Yam10}. It is neither embedded in the wind of an M giant companion nor a recurrent nova, making it representative of the majority of novae. In fact, V1723 Aql was the first northern classical nova bright enough for detailed observations in radio since the recent expansion of the VLA radio array.

Given a few reasonable assumptions, radio observations can be used to trace the bulk of the ejected mass of a nova simply and accurately. When we observe a nova in the radio regime, the spectrum reveals whether the emission is thermal bremsstrahlung radiation. A nova shell emitting thermal free-free emission starts as optically thick at all radio frequencies. During this stage of the nova's evolution, the radio flux densities are proportional to the surface area of the shell projected on the sky, and as the shell expands it becomes brighter at all radio frequencies. The flux density during this time should depend only on the distance to the nova, the temperature of the emitting material,  and the maximum velocity of the ejecta. However, as the density drops, radio frequency emission starts to penetrate through part of shell, resulting in a mix of optically thick and thin emission. During this time, the light curve peaks and starts to turn over. Once the photospheres have transitioned through the entire shell, we see purely optically thin emission. The shell transitions towards being optically thin at higher frequencies first, followed by the lower frequencies, until eventually the entire shell is optically thin at all wavelengths. The timing of the period when the light curve peaks and starts to turn over and the resulting rate of decline of the flux density light curve depends critically on the mass of the ejecta and on the density profile \citep{Bode08, Hjell79}. Therefore, the evolution of the radio opacity at different wavelengths can be used to infer density profiles of the ejected shell, and trace mass and density changes within the shell.

When modeling the behavior of a free-free emitting nova shell, we begin by making several additional assumptions. We assume a spherically symmetric nova shell, with a constant temperature of $10^4$ K throughout its evolution. We use a velocity distribution such that velocity is proportional to the distance to the white dwarf, known as the ``Hubble flow" velocity distribution, fitting the ratio between innermost and outermost velocity to the data \citep{Hjell79}. Finally, for the distribution of the ejected mass, we assume a smooth density profile that goes as $1/r^2$. This simple model has been relatively effective in explaining radio light curves of classical novae in the past, e.g. V1500 Cyg, QU Vul, and V723 Cas \citep{Hjell79, Tay88, Hey05}. However, only a small handful of classical novae have well-sampled radio light curves spanning the evolution from optically-thick rise to optically-thin decay. 

\section*{Observations and Results}

\begin{figure}[!ht]
\includegraphics[width=5.0in]{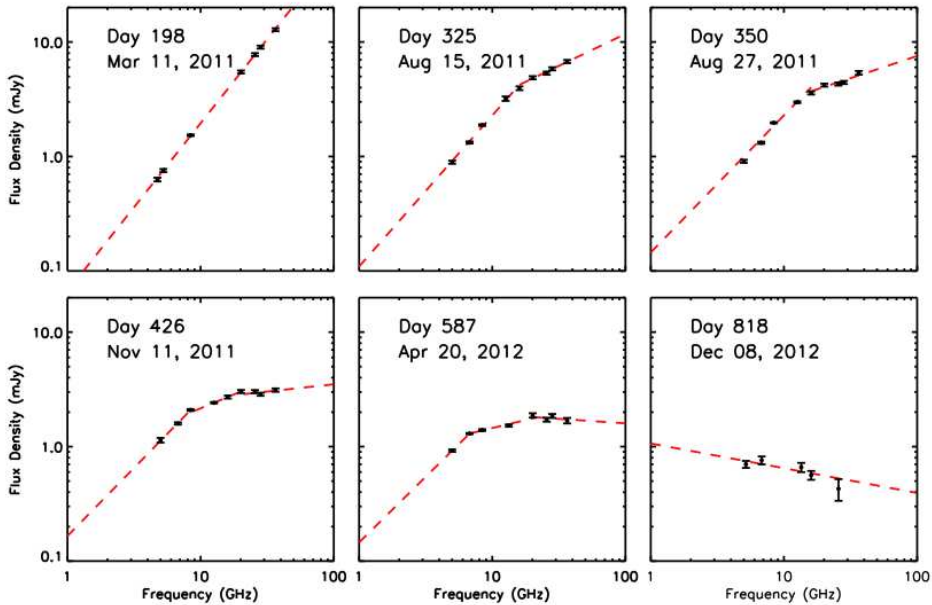}  
\caption{Spectrum of V1723 Aql during transition from optically thick to thin. The breaks in the spectrum reveal the transition of the radio photospheres from optically thick to optically thin.}
\label{fig:Breaks}
\end{figure}
We have now been monitoring this source with the VLA for over 2 years at multiple frequencies, starting 2 weeks after the initial discovery, and have produced one of the best, most detailed radio light curves of a classical nova to date. \citet{Krauss11} presented the initial report of the early part of the VLA observations. To determine the mass of the nova shell for V1723 Aql, we first consider the changes in the radio spectrum (Figure \ref{fig:Breaks}). On days 325 and 350 there is clearly one break in the spectrum, which has two parts: the rising optically thick portion at low frequencies, and a shallower spectrum revealing the region where material is transitioning between being optically thick and thin at higher frequencies. However, by day 426 there are two breaks visible. The lowest frequencies exhibit a spectrum of an optically thick photosphere, the middle frequencies show a mix of optically thin and thick emission, and the highest frequencies produce purely optically thin emission, demonstrating that the highest frequency photospheres have completely penetrated the nova shell.

The low frequency break shows the frequency and time the emitting shell has an optical depth of one ($\tau_\nu=1$), which gives us the emission measure at that point. The spectrum of the resulting transition region gives us a measurement of the density profile --- i.e. a spectral index of $\alpha \sim 0.6$ corresponds to a density profile of $1/r^2$ \citep{Bode08}. A $1/r^2$ density profile is at least initially consistent with our measurements, as we get a spectral index of 0.6 on day 325 between 20.1 and 36.5 GHz. The high frequency break shows that there is a distinct inner boundary to the shell, which radio emission can fully penetrate. We can find the outer radius at this epoch by using the expansion velocity and the time from outburst, and the inner radius by using the flux density to find the angular size of the $\tau=1$ photosphere at the second break. Combining this with the emission measure, $EM=\int_{R_i}^{R_o} n^2(r) dl$, we integrate density over the volume of the shell and get an independent mass estimate at each epoch we see the second break. Using the epochs where we see multiple breaks, we find an ejecta shell mass of $1.9 \pm 0.5 \times 10^{-4}$ $M_\odot$. Modeling our late time light curve, we find our best fit for an ejecta mass of $M \sim 2 \times 10^{-4}$ $M_\odot$, an ejecta temperature of $T\sim 10^4$ K, a maximum velocity v$_1 \sim 1500$ km/s, a distance of $d \sim 6.7$ kpc, and a ratio of minimum to maximum velocity v$_2/$v$_1 \sim 0.3$ 

\begin{figure}[!ht]
\includegraphics[width=5.0in]{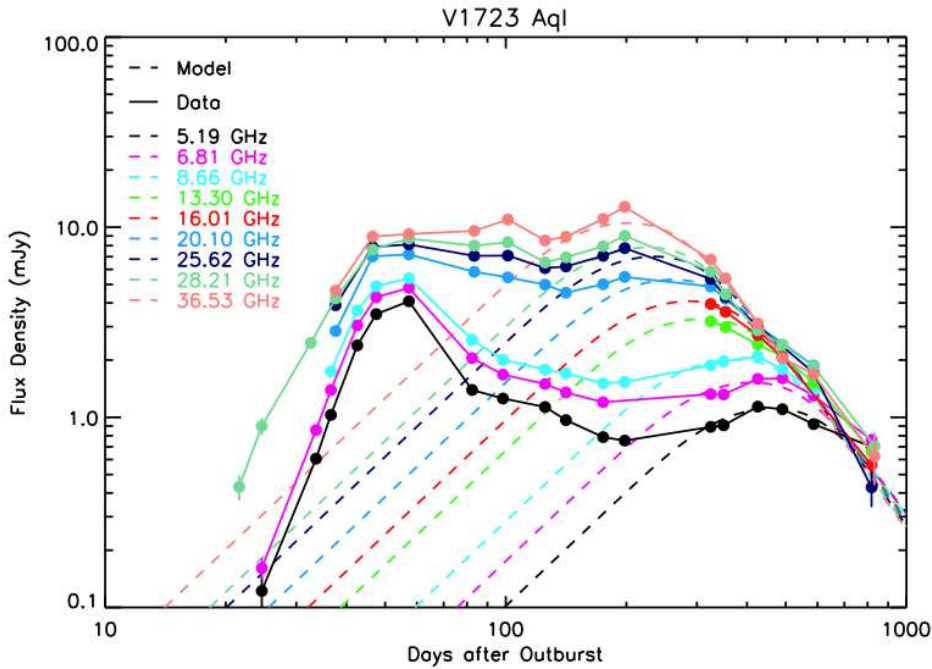} 
\caption{Full set of observations of V1723 Aql. Model indicates light curve of an expanding thermal shell of mass $M \sim 2 \times 10^{-4}$ $M_\odot$, ejecta temperature of $T\sim 10^4$ K, maximum velocity v$_1 \sim 1500$ km/s, distance of $d \sim 6.7$ kpc, and ratio of minimum to maximum velocity v$_2/$v$_1 \sim 0.3$.}
\label{fig:LC}
\end{figure}

If the ejection consisted of a single spherical impulsive ejection we would expect to see an optically thick spectrum that rises with frequency, with flux density increasing as $t^2$ as the brightness increases proportionally to the surface area of the shell during the first six months of a thermal shell's expansion. Contrasting this to our early observations of V1723 Aql, we instead see a distinct bump that is inexplicable by an expanding thermal shell alone (Figure \ref{fig:LC}), as discussed in \citet{Krauss11}. The radio flux density initially rises rapidly, proportional to $t^{3.3}$, then decreases and rises again, causing a dip in the light curve. In addition, the spectrum during the bump and the dip is not consistent with optically thick emission. During the initial rise the spectrum is shallower than is consistent with a purely optical thick expansion. Over the course of the bump, the source produces an optically thin flash and the spectrum flattens further, before returning to an optically thick spectrum when the flux density dips. These spectral changes are the opposite of what one would expect for an expanding optically thick source. 

While the light curve and spectrum are not consistent with an expanding shell alone, they could be explained by having multiple emission regions. If there were a shocked outside region then the increase in temperature in the shock-heated plasma would cause the post-shock ejecta to become optically thin at higher densities, resulting in the optically thin flash. This optically thin emission would  fade as the plasma cooled. The expansion of the bulk of the ejecta would then cause the second peak in the light curve as the underlying optically thick material expanded. Thus, instead of a single component model, we instead have a two component spectrum: the bulk of the material comprising of an expanding shell exhibiting optically thick free-free emission, and outside of that, a region that has been heated by shocks producing optically thin emission. Observations with Swift during the radio bump period detected V1723 Aql as an X-ray source \citep{Krauss11}; this X-ray emission also suggests strong shocks, supporting the two component model. Any secondary emission component must be outside of the optically thick shell, since any internal shock would not be visible through the optically thick material. Therefore, either the nova shell expanded into some outside material, or the ejecta itself may have collided with slower material that was ejected prior to this point. V1723 Aql is not embedded within the wind of a red giant companion, so there is no signature of a surrounding dense environment which might be the cause of strong shocks and synchrotron emission, unlike the cases of RS Oph and V407 Cyg \citep{OBrien06, Rupen08, Sok08, Chomiuk12}. This fact could indicate that the unusual light curve of V1723 Aql is a result of multiple periods of ejection, similar to the case of T Pyx in 2011 \citep{Netal2013}.

\section*{Final Thoughts}

\begin{figure}[!ht]
\includegraphics[width=4.8in]{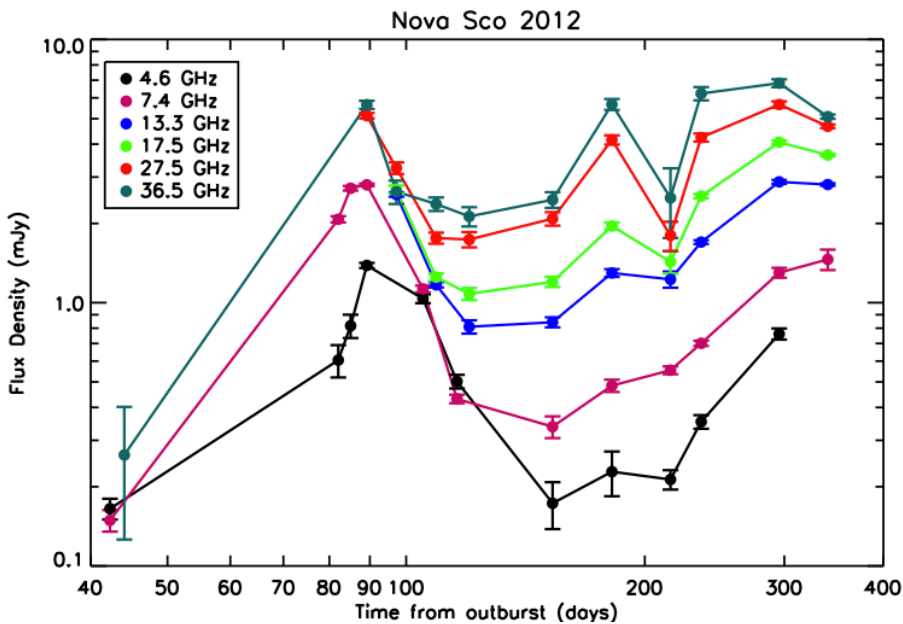} 
\caption{Radio Observations of Nova Sco 2012}
\label{fig:Sco}
\end{figure}

Prior to spring 2012, the bump in the light curve of V1723 Aql was a unique occurrence in radio observations of a classical novae, although hints of a similar phenomenon might have been precent in V1974 Cyg \citep{Lloyd96}. However, VLA observations of Nova Sco 2012 \citep{ATel4288} show a similar steep rise in the light curve, followed by a period where the flux density falls, then rises again (figure \ref{fig:Sco}). While V1723 Aql appears to be an otherwise typical classical nova, Nova Sco 2012 is not -- it was detected in $\gamma$-rays over the course of its outburst \citep{Cheung12}. One can speculate that whatever caused $\gamma$-rays in Nova Sco may be related to the cause of the bump in V1723 Aql. V1723 Aql is by all accounts a typical classical nova in the optical spectrum --- it is only in the radio regime that this irregularity reveals itself. Without early and frequent observations in the radio such a bump could be missed entirely, resulting in an incorrectly modeled light curve, which in turn affects distance and mass estimates. In fact, in many historical novae observed in radio prior to this the time and frequency coverage was insufficient to catch any hypothetical bump during early times, though this is not the case in particularly well sampled novae such as V1974 Cyg or FH Ser 1970 \citep{Hjell96}. It is possible that these early bumps are much more common than we realize.

\acknowledgements We are grateful to H. Uthas and J. Patterson for their advice and expertise. We thank NRAO for its generous allocation of time which made this work possible. The National Radio Astronomy Observatory is a facility of the National Science Foundation operated under cooperative agreement by Associated Universities, Inc.  J. Weston and J.L. Sokoloski acknowledge support from NSF award AST-1211778.

\bibliography{weston}

\end{document}